\documentstyle[12pt,psfig]{article}

\begin{document}         
\title{Dynamical phase transition in a neural network model with
noise: an exact solution} 
\author{Cristi\'an Huepe and 
        Maximino Aldana-Gonz\'alez\footnote{ Corresponding
        author. e-mail: maximino@control.uchicago.edu; tel: (773)-702-0946;
	fax: (773)-702-5863}}

\date{}

\maketitle

\begin{center}     
\it 
James Franck Institute, University of Chicago, 5640 South Ellis
Avenue.  Chicago, IL 60637, USA
\end{center}

\begin{center}
\it Submitted to the Journal of Statistical Physics.\\
January, 2002.
\end{center}

\begin{abstract}
  The dynamical organization in the presence of noise of a Boolean
  neural network with random connections is analyzed. For low levels
  of noise, the system reaches a stationary state in which the
  majority of its elements acquire the same value. It is shown that,
  under very general conditions, there exists a critical value
  $\eta_c$ of the noise, below which the network remains organized and
  above which it behaves randomly.  The existence and nature of the
  phase transition are computed analytically, showing that the
  critical exponent is $1/2$.  The dependence of $\eta_c$ on the
  parameters of the network is obtained. These results are then
  compared with two numerical realizations of the network.
\end{abstract}

\begin{flushleft}
{\bf Running title:} Dynamical phase transition in a neural network 
with noise.
\end{flushleft}

\pagebreak

\section{Introduction}

Boolean networks have been used to describe a wide variety of complex
systems.  They provide a common language for models of associative
memory \cite{91.4,89.2,82.1}, spin glasses
\cite{97.8,87.8,87.10,86.2}, dynamics of evolution
\cite{01.6,93.1,69.1}, and cellular automata
\cite{97.6,89.6,88.11,83.1}.  A typical Boolean network model consists
of a set of binary elements (also called nodes, neurons or spins,
depending on the context) which are connected among them to form a
net.  The value of each element at a given time depends on the value
at the previous time step of all the nodes that are connected to it.

The use of common tools of statistical physics has revealed a strong
parallel between Boolean networks and dynamical systems.  In this
context, several authors have studied the non-equilibrium dynamics of
deterministic Boolean networks and, in particular, the one of
\emph{neural networks} \cite{94.5,88.13}. Their work has unveiled the
existence of a variety of possible collective behaviors such as
synchronized oscillations or chaos \cite{96.7,90.1,82.1}.  On a
similar perspective, the influence of noise on the dynamics of Boolean
networks has been analyzed in \cite{89.3,89.5}.  It is shown that, for
increasing noise level, the barriers separating different attractors
decrease and then disappear.

We are interested in considering the changes in the dynamical
properties of a deterministic system in the presence of
noise. Following this motivation, we study a simple Boolean network
model exhibiting self-organization and analyze its tolerance to the
effect of noise.  We show analytically that the system undergoes a
dynamical second-order phase transition as its amount of randomness is
increased.

The paper is organized as follows. In section \ref{sec:def}, we
describe the neural network model with noise.  Section
\ref{sec:NumPhT} introduces the definition of the order parameter
characterizing the network and presents numerical evidence for the
phase transition by considering two particular network examples.  In
section \ref{sec:AnalPhT} we find the dynamical equation satisfied by
the order parameter and compute analytically its fixed points,
exhibiting the phase transition.  In section \ref{sec:2expls} we apply
these results to the two cases studied in section
\ref{sec:NumPhT}.  Finally, section \ref{sec:Conclusion} is our
conclusion.

\section{Definition of the model}               \label{sec:def}

%
\begin{figure}[t]
\centering
\psfig{file=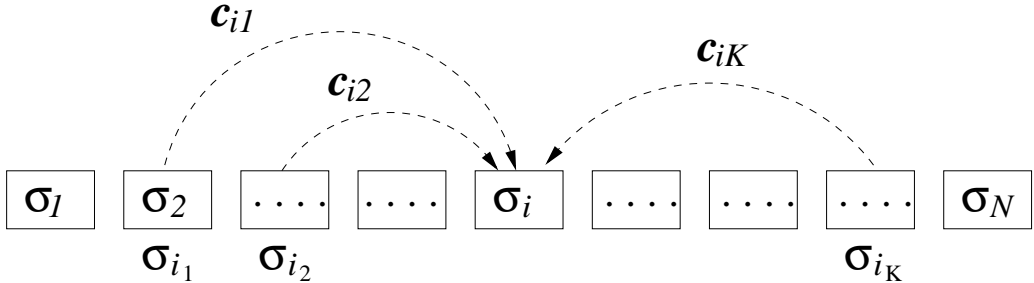,width=\hsize,clip=}
\caption[]{\small Schematic representation of the structure of the
 network. Every element $\sigma_i$ is connected to $K$ other elements
 $\{\sigma_{i_1},\dots,\sigma_{i_K}\}$ (the \emph{linkages}), which
 are chosen at random from the entire set $\{\sigma_i\}_{i=1\dots N}$.
 The contribution of each $\sigma_{i_j}$ element to the input function
 of $\sigma_i$ is weighted by $c_{ij}$ as given in equation
 (\ref{majority_rule}).}
\label{modelo}
\end{figure}
%

Consider a neural network composed of $N$ elements $\{\sigma_1,
\sigma_2,\dots,\sigma_N\}$, each of which can only take the values
$\sigma_i=-1$ or $\sigma_i=+1$.  Every $\sigma_i$ is randomly
connected to any $K$ elements of the network, which define its set of
\emph{linkages} $\{\sigma_{i_j}\}_{j=1,\dots,K}$ (see
Fig.\ref{modelo}).  The parameter $K$ is the \emph{connectivity} of
the network.  Each linkage $\sigma_{i_j}$ is weighted by an
independent random variable $c_{ij}$ that is chosen with a probability
density function (PDF) given by ${\mathrm P}_c(x)$.  The $N K$
connections of a network, and its corresponding weights, remain fixed
throughout the evolution of the system.

At every discrete time step, each $\sigma_{i}$ receives a signal
$+1$ or $-1$ equal to the \emph{input function}
\begin{equation}
f\left(c_{i1},\dots,c_{iK};
        \sigma_{i_1},\dots,\sigma_{i_K}\right)=
\mbox{Sign}\left\{\sum_{j=1}^{K} c_{ij} \sigma_{i_j}\right\}.
\label{majority_rule}
\end{equation}
For the particular case in which we have for all weights $c_{ij}=1$,
this definition corresponds to the \emph{majority rule}, in which $f$
takes the same value as the majority of the linkages.

Using the input function (\ref{majority_rule}), we define a stochastic
evolution rule for every $\sigma_{i}$ by introducing a noise intensity
$\eta$ such that
\begin{equation}
\sigma_i(t+1)=\left\{
\begin{array}{ccc}
f\left(c_{i1}\dots c_{iK};\sigma_{i_1}(t)\dots\sigma_{i_K}(t) \, \right)
&\mbox{with probability}& 1-\eta \\
& & \\
-f\left(c_{i1}\dots c_{iK};\sigma_{i_1}(t)\dots\sigma_{i_K}(t) \, \right)
&\mbox{with probability}& \eta .\\
\end{array}\right.
\label{rule:dynamics}
\end{equation}
The dynamics can thus be set from purely deterministic to purely
random by varying $\eta$ between $0$ and $1/2$.  Note that in the case
with $\eta=0$, the system will typically converge to an ordered state
in which all the $\sigma_i$ are equal.

Due to the presence of noise and to the randomness in the initial
assignment of the linkages, the statistical properties of the dynamics
of the network do not change if the connection weights or the linkages
are either time-independent or if they are randomly re-assigned at
every time step.  Using the language of boolean networks, this means
that for the model presented here the \emph{annealed} and
\emph{quenched} dynamics are equivalent \cite{86.3}.

\section{Numerical evidence}
                                                \label{sec:NumPhT}

In this section we perform a numerical study of the evolution of the
neural network model introduced above. We show that the system
undergoes a dynamical phase transition (for $N \rightarrow \infty$)
from an ordered to a disordered state as the noise intensity $\eta$ is
increased. The analytical expression for this transition will be
deduced in section \ref{sec:AnalPhT}.

%
\begin{figure}[h]
\centering
\psfig{file=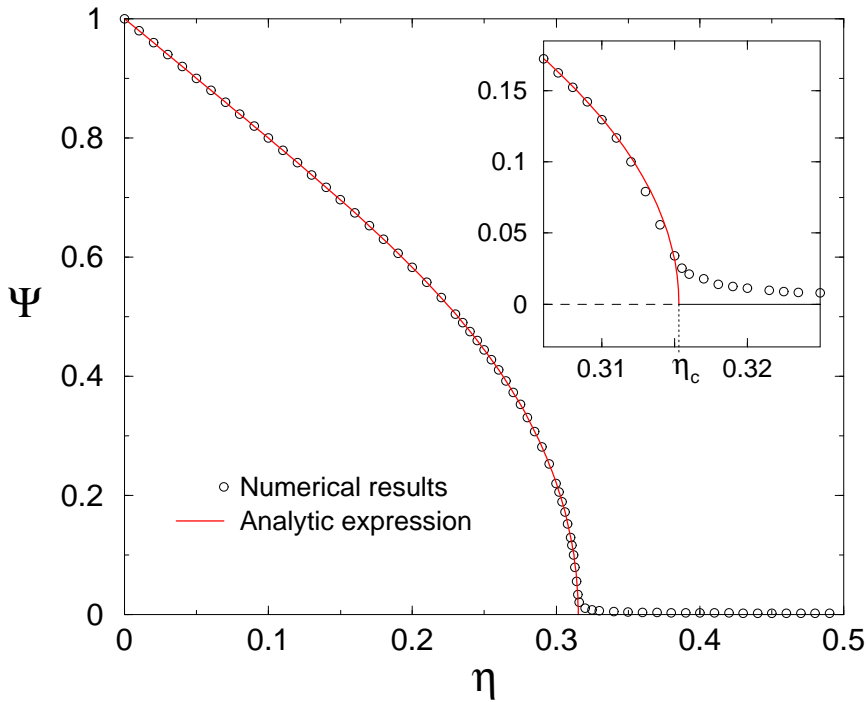,width=4.5in,clip=}
\caption[]{\small Bifurcation diagram of the order parameter $\Psi$ as
  a function of the noise intensity $\eta$ for a neural network model
  in which $c_{ij}=1$ for all weights.  A phase transition occurs at
  $\eta_c \simeq 0.3153$.  The numeric and analytic results only
  differ at $\eta \sim \eta_c$, where the blowup shows a slight
  difference due to finite size effects.}
\label{fig:symKonst}
\end{figure}
%

Let us define an order parameter that adequately describes the degree
of alignment of the elements of the network.  We first introduce
\begin{equation}                        \label{def:s(t)}
s(t) = \lim_{N \rightarrow \infty} \left[
        \frac{1}{N} \sum_{i=1}^{N} \sigma_i(t) \right].
\end{equation}
With this definition, $|s(t)| \approx 1$ for an ``ordered'' system in
which most elements take the same value, while $|s(t)| \approx 0$ for
a ``disordered'' system in which the elements randomly take values
$+1$ or $-1$.  For systems where the time-average of $|s(t)|$
converges, a time-independent order parameter can now be defined as
\begin{equation}
\Psi = \lim_{T \rightarrow \infty} 
        \frac{1}{T-T_0} \int_{T_0}^T |s(t)| dt,
\end{equation}
where $T_0$ can take any arbitrary finite value without changing
$\Psi$.

We computed numerically the evolution of the model for a network with
$N=100000$ elements, connectivity $K=11$, and random initial
conditions.  In practice, the order parameter $\Psi$ was obtained by
integrating $|s(t)|$ from $T_0=1000$ (to drop the initial relaxation
dynamics) until $T=10000$. A change to a larger integration time
produces negligible variations on the result.

%
\begin{figure}[t]
\centering
\psfig{file=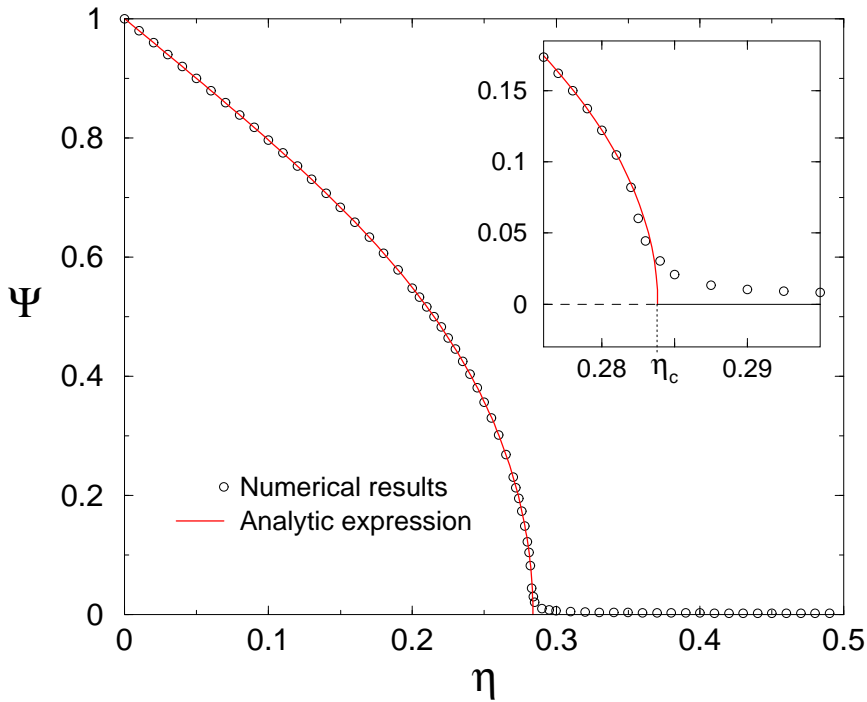,width=4.5in,clip=}
\caption{\small Bifurcation diagram of the order parameter $\Psi$ as
  a function of the noise intensity $\eta$ for a neural network model
  where the weights $c_{ij}$ follow the PDF defined in
  (\ref{eq:PDF_2}).  A phase transition occurs at $\eta_c \simeq
  0.2838$.  The blowup shows again the differences between the numeric
  and analytic results due to finite size effects (see Fig.
  \ref{fig:symKonst}).}
\label{fig:symUnif}
\end{figure}
%

The numerical results presented on Figure \ref{fig:symKonst} show the
bifurcation diagram of $\Psi$ as a function of the control parameter
$\eta$ for the case with $c_{ij}=1$, in which all connection weights
are equal. The input function (\ref{majority_rule}) then simply
becomes the majority rule.  It is apparent that the system undergoes a
phase transition at $\eta_c \approx 0.32$.  For $\eta < \eta_c$, all
elements in the system will tend to align either to $+1$ or to $-1$.
For $\eta > \eta_c$, their values are randomly distributed.  The
blowup on figure \ref{fig:symKonst} shows the usual finite size effect
on the phase transition, which smoothes the curve near $\eta_c$.

On figure \ref{fig:symUnif} we present the results for the case of a
network with fixed connection weights $c_{ij}$ that follow the PDF
\begin{equation}
{\mathrm P}_c(x) =\left\{
\begin{array}{cc}
1 & {\mathrm{if }} \, \, 0 \leq x \leq 1 \\
0 & {\mathrm{otherwise}}.\\
\end{array} \right.
\label{eq:PDF_2}
\end{equation}
The phase transition on this system is qualitatively equivalent to the
previous one, but the critical noise value is now changed to $\eta_c
\approx 0.28$.  The blowup shows again the finite size effects.

\section{Analytic solution}                                                    \label{sec:AnalPhT}

We will compute here the exact analytic expression that relates the
noise intensity control parameter $\eta$ and the order parameter
$\Psi$ for any given PDF of the connection weights ${\mathrm P}_c(x)$.

\subsection{Dynamics of the order parameter}
\label{sec:OPD}

First, we will relate the probability distribution of the system at a
time $t+1$ with the one at a time $t$. In order to do so, let us
define $\phi_N(t)$ as the fraction of elements in the network whose
value is $+1$ at time $t$:
\begin{equation}                                \label{def:phi(t)}
\phi_N(t) = 
        \frac{1}{N} \sum_{i=1}^N \frac{\sigma_i(t) + 1}{2}.
\end{equation}
In the thermodynamic limit $N\rightarrow\infty$, the above quantity
transforms into the probability that at time $t$ any arbitrary node 
$\sigma_i$ acquires the value $+1$:
\begin{equation}
\phi(t)\equiv {\mathbf P}_t\left\{\sigma_i = +1\right\}  = 
\lim_{N\rightarrow\infty}\phi_N(t)
\end{equation}
Note that from definition (\ref{def:s(t)}), the relation between
$s(t)$ and $\phi(t)$ is simply given by 
\begin{equation}
s(t) = 2 \phi(t) - 1.
\label{s_t}
\end{equation}
Thus, in a fully ordered state we have $|s(t)| = 1$ and $\phi(t) = 0$
or $1$, while in a fully disordered state we have $|s(t)| = 0$ and
$\phi(t) = 1/2$.

It is useful to define $\xi_i(t)$ as the argument of the Sign function
appearing in the definition of $f$ for the $i$-th element at a time
$t$ (see equation (\ref{majority_rule})),
\begin{equation}
\xi_i(t) = \sum_{j=1}^K c_{ij} \sigma_{i_j}(t).
\end{equation}
If the linkages of every node are assigned in a sufficiently random
way\footnote{If the linkages are not assigned randomly, the situation
  changes. For example if they are chosen among the first neighbors of
  every element, there is no phase transition and the analysis
  presented here is not applicable.}, the products
$c_{ij}\sigma_{i_j}(t)$ can be considered as independent random
variables. Therefore, if we denote by ${\mathrm{P}}_{\xi(t)}(x)$ and
${\mathrm{P}}_{c\sigma(t)}(x)$ the PDF associated to $\xi_i$ and to the
product $c_{ij}\sigma_{i_j}$ respectively, then
${\mathrm{P}}_{\xi(t)}(x)$ is simply given by the $K$-fold convolution
of ${\mathrm{P}}_{c\sigma(t)}(x)$ with itself:

\begin{equation}
{\mathrm{P}}_{\xi(t)}(x) = \underbrace{{\mathrm{P}}_{c\sigma(t)}\ast 
{\mathrm{P}}_{c\sigma(t)}\ast \cdots
\ast {\mathrm{P}}_{c\sigma(t)}(x)}_{K\ \mbox{times}}.
\label{convolution}
\end{equation}
In terms of ${\mathrm{P}}_{\xi(t)}(x)$, the probability $I(t)$ of having
the input function $f = +1$ at time $t$ can be computed as
\begin{equation}                        \label{def:I}
I(t) \equiv {\mathbf P}_t\left\{f=+1\right\}= 
\int_0^\infty {\mathrm{P}}_{\xi(t)}(x) dx.
\end{equation}
Using the updating rule (\ref{rule:dynamics}), we can now directly
write the probability $\phi(t+1)$ of having
$\sigma_i(t+1) = +1$ in terms of $I(t)$ and $\eta$:
\begin{equation}                        \label{eq:FundEq}
\phi(t+1) =
        I(t) \left[1 - \eta \right] + \left[1 - I(t) \right] \eta.
\end{equation}
This master equation describes the stochastic dynamics of the network.
Its fixed points as a function of $\eta$ will generate the bifurcation
diagram showing the phase transition.

\subsection{Probability distribution of the input function}
\label{sec:PDF}

In order to express equation (\ref{eq:FundEq}) in a closed form, we
must find how $I(t)$ relates to $\phi(t)$.  For this, we first compute
${\mathrm{P}}_{\xi(t)}(x)$ which is in turn determined by
${\mathrm{P}}_{c\sigma(t)}(x)$. In Fourier space, the convolution
appearing in (\ref{convolution}) acquires the simple form
\begin{equation}
\hat{{\mathrm{P}}}_{\xi(t)}(\lambda) = 
\left[\hat{{\mathrm{P}}}_{c \sigma(t)}(x) \right]^K,
\label{P_xi1}
\end{equation}
where $\hat{{\mathrm{P}}}_{\xi(t)}(\lambda)$ and
$\hat{{\mathrm{P}}}_{c \sigma(t)}(\lambda)$ are the Fourier transforms
of ${\mathrm{P}}_{\xi(t)}(x)$ and ${\mathrm{P}}_{c \sigma (t)}(x)$
respectively.

Since the connection weights $c_{ij}$ are distributed according to the
probability function ${\mathrm{P}}_c(x)$, and the variables
$\sigma_i(t)$ evaluate to $+1$ with probability $\phi(t)$ and to $-1$
with probability $1-\phi(t)$, it follows that the PDF of the products
$c_{ij}\sigma_{i_j}$ is given by
\begin{equation}                        \label{eq:pbb_xi}
{\mathrm{P}}_{c \sigma (t)}(x) =
        \phi(t) {\mathrm{P}}_c(x) + 
        \left[ 1 - \phi(t) \right] {\mathrm{P}}_c(-x).
\end{equation}
Taking the Fourier transform of the previous expression, and inserting
the result into equation (\ref{P_xi1}), one gets
\begin{equation}
\hat{{\mathrm{P}}}_{\xi(t)}(\lambda) = 
\left[\hat{{\mathrm{P}}}_{c}^*(\lambda) +
\left(\hat{{\mathrm{P}}}_{c}(\lambda) -
\hat{{\mathrm{P}}}_{c}^*(\lambda)\right)\phi(t)\right]^K,
\label{P_xi2}
\end{equation}
where the `${}^*$' denotes complex conjugation. From the above
expression it is apparent that ${\mathrm{P}}_{\xi(t)}(x)$, and
consequently $I(t)$, are polynomial functions of $\phi(t)$. Therefore,
equation (\ref{eq:FundEq}) is a polynomial of degree $K$ in $\phi(t)$,
with solutions that depend on the value of the noise intensity
$\eta$. As we will see later, the roots of this polynomial will
furnish the bifurcation diagram of the order parameter.

For convenience, we will write our results in terms of $s(t)$ instead
of $\phi(t)$ (see equation (\ref{s_t})).  Substituting $\phi(t)=
\left[s(t)+1\right]/2$ in expression (\ref{P_xi2}) we get
\begin{equation}
\hat{{\mathrm{P}}}_{\xi(t)}(\lambda) = 
\left[\frac{\hat{{\mathrm{P}}}_c(\lambda) + 
    \hat{{\mathrm{P}}}_c^*(\lambda)}{2} +
  \frac{\hat{{\mathrm{P}}}_c(\lambda) - 
    \hat{{\mathrm{P}}}_c^*(\lambda)}{2} s(t)\right]^K.
\label{P_xi3}
\end{equation}
By denoting  $\hat{g}(\lambda)$ and $\hat{h}(\lambda)$, the real and
imaginary parts of $\hat{{\mathrm{P}}}_c(\lambda)$ respectively,
$\hat{{\mathrm{P}}}_{\xi(t)}(\lambda)$ can then be written as
\begin{eqnarray}
\hat{{\mathrm{P}}}_{\xi(t)}(\lambda) &=& 
\left[\hat{g}(\lambda)+i\hat{h}(\lambda)s(t)\right]^K \nonumber \\
 &=&\sum_{m=0}^K {K\choose m}\left[\hat{g}(\lambda)\right]^{K-m}\nonumber
\left[i\hat{h}(\lambda)s(t)\right]^m,
\end{eqnarray}
whose inverse Fourier transform is
\begin{equation}
{\mathrm{P}}_{\xi(t)}(x) = 
\sum_{m=0}^K
\left\{
\frac{i^m}{2\pi}{K\choose m}
\int_{-\infty}^{\infty}
\left[\hat{g}(\lambda)\right]^{K-m}
\left[\hat{h}(\lambda)\right]^m e^{-i\lambda x}dx
\right\}
\left[s(t)\right]^m.
\label{eq:Pxi}
\end{equation}
We are now in position of computing $I(t)$. Using equations
(\ref{def:I}) and (\ref{eq:Pxi}), and moving the integrals inside
the sum we have
\begin{equation}
I(t) = \sum_{m=0}^K a_m \left[ s(t) \right]^m,
\end{equation}
where the $a_m$ are constant coefficients that depend only on
$\mathrm{P}_c(\lambda)$ and are given by
\begin{equation}
a_m = \frac{i^m}{2 \pi}
        {K\choose m}
        \int_{0}^{\infty} \int_{-\infty}^{\infty}
                        \left[ \hat{g}(\lambda) \right]^{K-m}
                        \left[ \hat{h}(\lambda) \right]^m
                        e^{- i \lambda x} \, d\lambda \, dx.
\end{equation}
We can formally integrate over $x$ by replacing $e^{-i \lambda x}
\rightarrow e^{-(i \lambda + \epsilon) x}$, computing the new
$x$-integral, and then evaluating the result at $\epsilon = 0$.  Our
final formula for the $a_m$ coefficients gives
\begin{equation}                        \label{eq:a_m_Final}
a_m = \frac{- i^{m+1}}{2 \pi} 
        {K\choose m}
        \int_{-\infty}^{\infty}
                        \frac{1}{\lambda}
                        \left[ \hat{g}(\lambda) \right]^{K-m}
                        \left[ \hat{h}(\lambda) \right]^m
                                                d\lambda.
\end{equation}
Note that $a_m = 0$ for all even values of $m$.
Indeed, since the function $\hat{g}(\lambda)$ is even and $\hat{h}(\lambda)$
is odd, the integrand will be antisymmetric for any even $m$,
thus vanishing the integral.

%
\begin{figure}
{\centering
\psfig{file=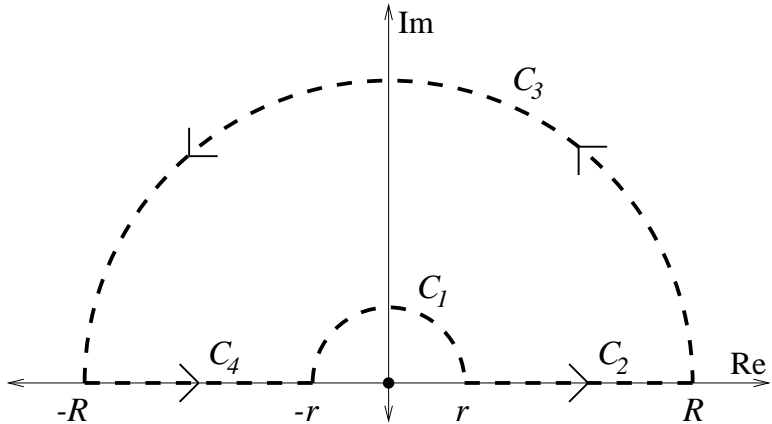,width=4in,clip=}}
\caption[]{\small Integration contour ${\mathcal{C}}$ on the complex 
plane that was used to compute (\ref{eq:z}).  The pole at the origin
is circumvented by the half-circle ${\mathcal{C}}_1$ of radius $r$. }
\label{fig:Path}
\end{figure}
%

It will be useful for later calculations to compute the value of
$a_0$.  In fact, it turns out that $a_0=1/2$ for every $K$ and any
``well behaved'' function $\mathrm{P}_c(x)$. This can be readily
proven by extending the integral in (\ref{eq:a_m_Final}) to the
complex plane.  The integration path ${\mathcal{C}}$ shown on figure
\ref{fig:Path} contains no poles, therefore
\begin{equation}                \label{eq:z}
\oint_{\mathcal{C}} \frac{1}{z} \left[ \hat{g}(z) \right]^K dz = 0.
\end{equation}
Since ${\mathrm{P}}_c(x)$ is a PDF, Parseval's theorem guarantees that
its Fourier transform is square integrable, which implies
$\int_{-\infty}^{\infty} \hat{g}^2(\lambda) d \lambda < \infty$.
Therefore, the contribution of segment ${\mathcal{C}}_3$ to the
integral in (\ref{eq:z}) is zero for $R \rightarrow \infty$.  On the
other hand, being $\hat{g}(\lambda)$ the real part of the Fourier
transform of a PDF, one has $\hat{g}(0) = 1$ and thus, for any well
behaved function it is possible to approximate $\hat{g}(r)\simeq 1$
over the small segment ${\mathcal C}_1$. This allows us to compute 
$\int_{{\mathcal C}_1} \frac{1}{z}
\left[ \hat{g}(z) \right]^K dz = \int_\pi^0 i d\theta = - i \pi$.
Replacing into (\ref{eq:z}), we obtain the value of the integral in
(\ref{eq:a_m_Final}) and find that $a_0 = 1/2$ for any
${\mathrm{P}}_c(x)$ and any $K$.

\subsection{Computing the bifurcation diagram}
\label{subsec:bifdiag}

We can now combine the main results of sections \ref{sec:OPD} and
\ref{sec:PDF} to find an analytic expression relating $\eta$ and
$\Psi$. We have shown that $I(t)$ is a polynomial of degree $K$ in
$s(t)$. Therefore, in terms of $s(t)$, the master equation
(\ref{eq:FundEq}) governing the dynamics of the network becomes
\begin{equation}
s(t+1) =
2\left(1-2\eta\right)\left(a_1s(t)+a_3[s(t)]^3+\cdots+
a_K[s(t)]^K\right).
\label{eq:master_s_t}
\end{equation} 
In the limit $t\rightarrow\infty$, $s(t)$ will asymptotically approach
a fixed point $s$ which, from the above equation, obeys
\begin{equation}
s =
2\left(1-2\eta\right)\left(a_1s+a_3s^3+\cdots+
a_Ks^K\right).
\label{eq:master_s}
\end{equation} 
It is important to point out that if the probability function
$\mathrm{P}_c(x)$ is symmetric, there is no phase transition. Indeed,
if $\mathrm{P}_c(x)$ satisfies $\mathrm{P}_c(x)=\mathrm{P}_c(-x)$,
then the imaginary part $\hat{h}(\lambda)$ of its Fourier transform
would be identically zero.  It follows from equation
(\ref{eq:a_m_Final}) that $a_m=0\ \forall\ m$ and, therefore, equations
(\ref{eq:master_s_t}) and (\ref{eq:master_s}) give the trivial result
$s=0$ as the only possible solution for the dynamics.

Equation (\ref{eq:master_s}) is always satisfied by $s=0$. However, as
$\eta$ is varied, this solution becomes unstable as other ones appear.
Discarding the solution $s=0$ and solving equation (\ref{eq:master_s})
for $\eta$, we get
\begin{equation}
\eta=\frac{a_1-1/2+a_3s^2+\cdots+a_Ks^{K-1}}
{2(a_1+a_3s^2+\cdots+a_Ks^{K-1})}.
\label{eq:division}
\end{equation}
By dividing the polynomials and neglecting the terms of order $s^4$
and higher in the resulting expression\footnote{As it can be seen in
figures \ref{fig:symKonst} and \ref{fig:symUnif}, the order parameter
vanishes continuously when passing from the ordered to the disordered
phase. Therefore, in the vicinity of the phase transition $s\approx
0$.}, we obtain
\begin{equation}
\eta-\eta_c=\frac{a_3}{4a_1^2}s^2
\label{eq:phasetrans}
\end{equation}
where $\eta_c$ is defined as
\begin{equation}
\eta_c= \frac{1}{2}\left(1 - \frac{1}{2 a_1} \right).
\label{eq:eta_c}
\end{equation}
Since $a_3<0$, equation (\ref{eq:phasetrans}) implies that real
non-zero solutions for $s$ in (\ref{eq:master_s}) only exist if
$\eta<\eta_c$.  For $\eta>\eta_c$ the solutions of equation
(\ref{eq:phasetrans}) are imaginary and therefore $s=0$ is the only
acceptable solution of (\ref{eq:master_s}). We thus conclude that a
phase transition with critical exponent $1/2$ will occur at $\eta =
\eta_c$.  The explicit behavior of the order parameter $\Psi=|s|$ near
the transition will be

\begin{equation}                        \label{eq:|s|}
\Psi = \left\{ 
\begin{array}{lcr}
\frac{2 a_1}{\sqrt{|a_3|}}
\left(\eta_c - \eta\right)^{1/2} & \mbox{for} &
\eta<\eta_c \\
 & & \\
0 & \mbox{for} & \eta>\eta_c
\end{array}\right.
\end{equation}
%

\section{Examples of the analytic solution}
                                                \label{sec:2expls}

%
\begin{figure}[t]
\centering
\psfig{file=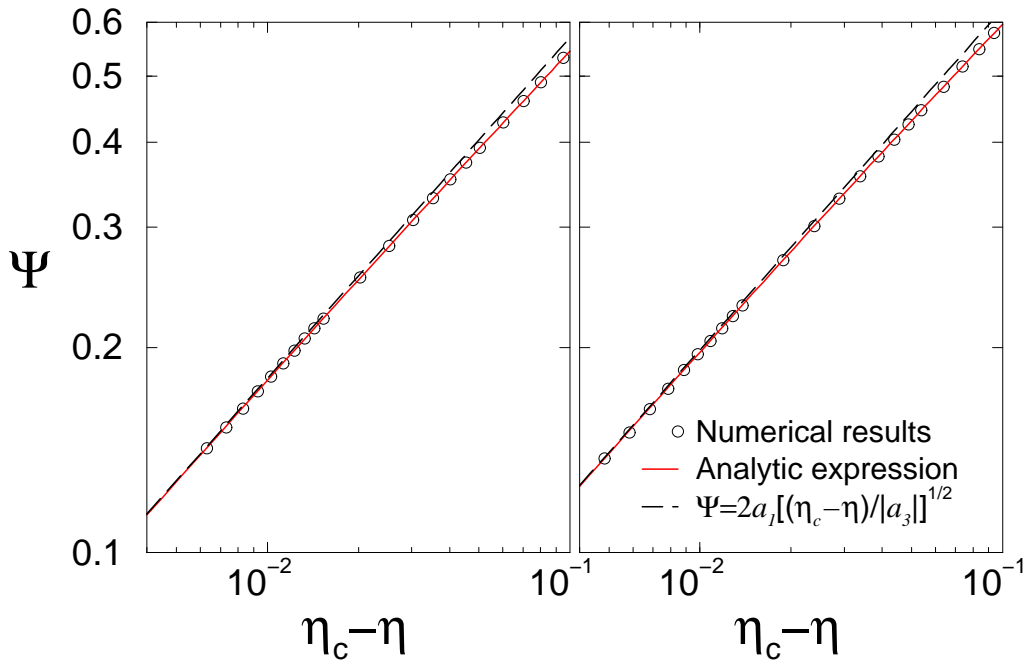,width=\hsize,clip=}
\caption[]{\small Log-log plot of the order parameter $\Psi$ as a 
function of the distance to the critical noise value $\eta_c-\eta$.
The left side graph corresponds to the case with equal connection
weights $c_{ij}=1$.  The right hand graph is for uniformly distributed
connection weights.  The values of $\eta_c = 0.3153$ (left) and
$\eta_c = 0.2838$ (right) were obtained from equation
(\ref{eq:eta_c}).  The numerical results (circles) and the analytic
solution (solid line) approach asymptotically the dashed line
representing the bifurcation form (\ref{eq:|s|}).}
\label{fig:psi_vs_eta}
\end{figure}
%

In this section we compare our analytic solution with the 
numerical results presented in section \ref{sec:NumPhT}.

We consider first the case with $c_{ij} = 1$ for all connections. The
PDF of these connection weights is then ${\mathrm{P}_c}(x) =
\delta(x-1)$, and we have
\begin{equation}
\hat{{\mathrm{P}}}_c(\lambda) =
        \int_{-\infty}^{\infty} \delta(x-1) e^{i \lambda x} dx =
        \cos(\lambda) + i \sin(\lambda).
\end{equation}
Therefore, the coefficients $a_m$ are given by
\begin{equation}
a_m=\frac{-i^{m+1}}{2\pi}{K\choose m}\int_{-\infty}^{\infty}
\frac{1}{\lambda}\left[\cos(\lambda)\right]^{K-m}\left[\sin(
\lambda)\right]^m d\lambda.
\label{eq:a_m_delta}
\end{equation}
For the other case, in which the connection weights are uniformly
distributed in the interval $[0,1]$ as given in expression 
(\ref{eq:PDF_2}), we get
\begin{equation}
\hat{{\mathrm{P}}}_c(\lambda) =
        \int_{0}^{1} e^{i \lambda x} dx =
                \frac{\sin(\lambda)}{\lambda} + 
                        i \frac{1 - \cos(\lambda)}{\lambda},
\end{equation}
and the corresponding coefficients $a_m$ are given by
\begin{equation}
a_m=\frac{-i^{m+1}}{2\pi}{K\choose m}\int_{-\infty}^{\infty}
\frac{1}{\lambda^{K+1}}\left[\sin(\lambda)\right]^{K-m}
\left[1-\cos(\lambda)\right]^m d\lambda.
\label{eq:a_m_uniform}
\end{equation}
We evaluated the integrals in (\ref{eq:a_m_delta}) and
(\ref{eq:a_m_uniform}) for the case $K=11$, which was studied in
section \ref{sec:NumPhT}.  By replacing them into equation
(\ref{eq:division}) we explicitly obtain $\eta$ as a function of $s$
for these two particular cases.  The solid curves on figures
\ref{fig:symKonst} and
\ref{fig:symUnif} show the analytic bifurcation diagram that is found
through this procedure. The agreement with the results our numerical
simulation is excellent, confirming our assumption that all elements
can be considered as independent random variables, even if the network
connections and weights are kept fixed throughout the evolution of the
system.

\begin{figure}[ht]
\centering
\psfig{file=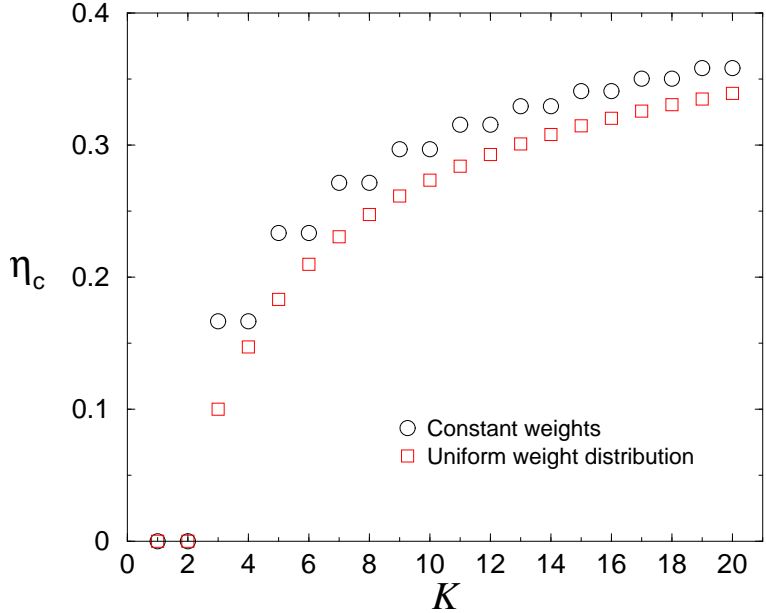,width=4in,clip=}
\caption[]{\small Critical noise level $\eta_c$ for various values 
of the connectivity $K$ and two different weight distributions:
$c_{ij}=1$ for all connections (circles) and $c_{ij}$ uniformly
distributed in $[0,1]$ (squares).  As the connectivity increases, the
amount of noise needed to disorganize the system must increase.  For
large $K$, the value of $\eta_c$ will tend asymptotically to the
maximum noise level $1/2$. For $K \leq 2$ both cases have $\eta_c=0$,
and the system becomes disorganized for any noise $\eta > 0$.  In the
constant weight case, the values of $\eta_c$ at consecutive odd and
even values of $K$ are equal (see text).}
\label{fig:etac_vs_K}
\end{figure}
%

Once the values of the coefficients $a_m$ are obtained, the critical
noise value $\eta_c$ can be readily calculated by using equation
(\ref{eq:eta_c}).  Figure \ref{fig:psi_vs_eta} shows a log-log graph of
the vanishing value of $\Psi$ as a function of $\eta_c-\eta$, both for
the constant connection weight case ($\eta_c \simeq 0.3153$) and for
the uniform distribution weight case ($\eta_c \simeq 0.2838$).  These
results are the same that were presented in section \ref{sec:NumPhT}.
On the displayed windows, finite-size effects around $\eta_c$ are not
visible. As can be seen on this figure, the numerical results coincide
perfectly with the analytic solution (\ref{eq:division}) (solid curve)
and with the asymptotic behavior given in (\ref{eq:|s|}) (dashed
line).

Figure \ref{fig:etac_vs_K} shows the critical noise level $\eta_c$ as
a function of the connectivity $K$ for the two particular cases
studied.  As the connectivity is increased, the phase transition
appears at higher levels of noise and it is increasingly hard to
disorder the system. As the system becomes more and more correlated,
$\eta_c$ will asymptotically approach its maximum value of $1/2$ (data
not shown).  For the case in which $c_{ij}=1$ for all connections, the
majority rule is not well defined for even values of $K$, since there
can be equal number of elements with $\sigma_i=+1$ and $\sigma_i=-1$.
This justifies the degeneracy observed for $\eta_c$ with respect to
consecutive odd and even values of $K$.  For $K=2$ the system will be
disorganized at any $\eta_c>0$ and no phase transition occurs.

\section{Conclusions}                   \label{sec:Conclusion}

We have shown the existence of a noise-driven phase transition in a
neural network with random connections. We found an exact analytic
solution of the stochastic equation which governs the dynamics of the
system. By finding its fixed points as a function of noise we
constructed the bifurcation diagram, which shows that the phase
transition is of second-order with a critical exponent of $1/2$.

Besides the randomness of the network connection, our work was carried
out with very few assumptions.  One of these was to consider the
connection weights as statistically independent variables.  This
condition, however, is not satisfied by many models of interest such
as the Hopfield neural networks \cite{82.1}.  It would therefore be of
great interest to generalize our approach to cases in which the
connection weights are statistically correlated.  Given that these
conditions are satisfied, the existence of the phase transition only
seems to require that the PDF of the connection weights
${\mathrm{P}}_c(x)$ is a non-symmetric but otherwise arbitrary
function, as shown on section \ref{subsec:bifdiag}.

The general framework under which the results on this paper were
derived leads us to believe that this type of noise-driven phase
transitions from an ordered to a disordered state must be a robust
feature of systems with elements that are somehow randomly connected
\cite{90.5}.

\section*{Acknowledgements}
We would like to thank Leo Kadanoff for his encouragement and valuable
scientific discussions. This work was supported in part by the MRSEC
Program of the National Science Foundation under award number 9808595,
and by the NSF DMR 0094569. M. Aldana also acknowledges
CONACyT-M\'exico for a postdoctoral grant and the Santa Fe Institute of
Complex Systems for partial support through the David and Lucile
Packard Foundation Program in the Study of Robustness.


\bibliographystyle{plain}
\bibliography{bibliography}

\begin{thebibliography}{10}

\bibitem{96.7}
H.~D.~I. Abarbanel, M.~I. Rabinovich, A.~Selverston, and M.~V. Bazhenov.
\newblock Synchronization in neural networks.
\newblock {\em Physics-Uspeki}, 39(4):337--362, 1996.

\bibitem{94.5}
B.~Cheng and D.~M. Titterington.
\newblock Neural networks: a review from a statistical perspective.
\newblock {\em Statistical Science}, 9(1):2--54, Feb 1994.

\bibitem{97.6}
J.~A. De~Sales, M.~L. Martins, and D.~A. Stariolo.
\newblock Cellular automata model for gene networks.
\newblock {\em Physical Review E}, 55(3):3262--3270, Mar 1997.

\bibitem{87.8}
B.~Derrida.
\newblock Dynamical phase transitions in non-symmetric spin glasses.
\newblock {\em Journal of Physics A: Mathematical and General}, 20:L721--L725,
  1987.

\bibitem{86.2}
B.~Derrida and H.~Flyvbjerg.
\newblock Multivalley structure in kauffman model - analogy with spin-glasses.
\newblock {\em Journal of Physics A: Mathematical and General},
  19(16):1003--1008, Nov 1986.

\bibitem{86.3}
B.~Derrida and Y.~Pomeau.
\newblock Random networks of automata - a simple annealed approximation.
\newblock {\em Europhysics Letters}, 1(2):45--49, Jan 1986.

\bibitem{90.5}
J.~Doyne Farmer.
\newblock A roseta stone for connectionism.
\newblock {\em Physica D}, 42(1-3):153--187, Jun 1990.

\bibitem{89.6}
H.~Flyvbjerg.
\newblock Recent results for random networks of automata.
\newblock {\em Acta Physica Polonica B}, 20(4):321--349, Apr 1989.

\bibitem{89.5}
O.~Golinelli and B.~Derrida.
\newblock Barrier heights in the kauffman model.
\newblock {\em Journal De Physique}, 50(13):1587--1601, Jul 1989.

\bibitem{82.1}
J.~J. Hopfield.
\newblock Neural networks and physical systems with emergent collective
  computational abilities.
\newblock {\em Proceedings of the National Academy of Sciences},
  79(8):2554--2558, Apr 1982.

\bibitem{69.1}
Stuart~A. Kauffman.
\newblock Metabolic stability and epigenesis in randomly constructed nets.
\newblock {\em Journal of Theoretical Biology}, 22:437--467, 1969.

\bibitem{93.1}
Stuart~A. Kauffman.
\newblock {\em The Origins of Order: Self-Organization and Selection in
  Evolution}.
\newblock Oxford University Press, Oxford, 1993.

\bibitem{88.11}
K.~E. K{\"{u}}rten.
\newblock Correspondence between neural threshold networks and kauffman boolean
  cellular automata.
\newblock {\em Journal of Physics A: Mathematical and General},
  21(11):L615--L619, Jun 1988.

\bibitem{88.13}
K.~E. K{\"{u}}rten.
\newblock Critical phenomena in model neural netwoks.
\newblock {\em Physics Letters A}, 129(3):157--160, May 1988.

\bibitem{87.10}
M.~Mezard, G.~Parisi, and M.~A. Virasoro.
\newblock {\em Spin Glass Theory and Beyond}.
\newblock World Scientific, Singapore, 1987.

\bibitem{89.3}
E.~N. Miranda and N.~Parga.
\newblock Noise effects in the kauffman model.
\newblock {\em Europhysics Letter}, 10(4):293--298, Oct 1989.

\bibitem{97.8}
D.~Petters.
\newblock Patch algorithms in spin glasses.
\newblock {\em International Journal of Modern Physics C}, 8(3):595--600, Jun
  1997.

\bibitem{91.4}
K.~Sakai and Y.~Miyashita.
\newblock Neural organization for the long-term-memory of paired associates.
\newblock {\em Nature}, 354(6349):152--155, Nov 1991.

\bibitem{89.2}
D.~Sherrington and K.~Y.~M. Wong.
\newblock Random boolean networks for autoassociative memory.
\newblock {\em Physics Reports: Review Section of Physics Letters},
  184(2-4):293--299, Dec 1989.

\bibitem{90.1}
Lipo Wang, Elgar~E. Pichler, and John Ross.
\newblock Oscillations and chaos in neural networks - an exactly solvable
  model.
\newblock {\em Proceedings of the National Academy of Sciences of the United
  States of America}, 87(23):9467--9471, Dec 1990.

\bibitem{01.6}
Claus~O. Wilke, Cristopher Ronnenwinkel, and Thomas Martinetz.
\newblock Dynamic fitness landscapes in molecular evolution.
\newblock {\em Physics Reports}, 349(5):395--446, Aug 2001.

\bibitem{83.1}
Stephen Wolfram.
\newblock Statistical mechanics of cellular automata.
\newblock {\em Reviews of Modern Physics}, 55(3):601--644, Jul 1983.

\end{thebibliography}

\end{document}